\documentclass[sigconf, nonacm, natbib=true]{acmart}
\usepackage{xspace}

\usepackage{multirow}   
\usepackage{booktabs}   
\usepackage{array}      
\usepackage{dblfloatfix} 

\usepackage{cleveref}

\usepackage{svg}

\usepackage{xcolor}

\newcolumntype{H}{>{\setbox0=\hbox\bgroup}r<{\egroup}@{}}

\newcommand{\lib}{RISE\xspace}
\newcommand{\dsii}{DS2I\xspace}
\newcommand{\pisa}{PISA\xspace}

\newcommand{\cw}{\textsc{CW09}\xspace}
\newcommand{\cc}{\textsc{CCNews}\xspace}
\newcommand{\ef}{\texttt{ef\xspace}}
\newcommand{\upef}{\texttt{upef\xspace}}
\newcommand{\opt}{\texttt{pef\xspace}}
\newcommand{\optcomp}{\texttt{pefcomp\xspace}}
\newcommand{\svb}{\texttt{svb\xspace}}
\newcommand{\bic}{\texttt{bic\xspace}}

\newcommand{\todo}[1]{\textcolor{red}{TODO: #1}}
\renewcommand{\todo}[1]{}

\acmConference[CIKM 2026]{The 35th ACM International Conference on Information and Knowledge Management}{November 7--11, 2026}{Rome, Italy}

\title{\lib: A Rust Library for Inverted Index Search Engines}

\author{Angelo Savino}
\orcid{0009-0002-0298-1546}
\affiliation{
  \institution{University of Pisa}
  \city{Pisa}
  \country{Italy}
}
\email{a.savino6@studenti.unipi.it}

\author{Rossano Venturini}
\orcid{0000-0002-9830-3936}
\affiliation{
  \institution{University of Pisa}
  \city{Pisa}
  \country{Italy}
}
\email{rossano.venturini@unipi.it}

\begin{document}

\begin{abstract}
Inverted indexes are a crucial data structure for efficient information retrieval in large text corpora. 
They enable fast full-text search by mapping each term to the documents in which it appears, on top of which efficient algorithms quickly retrieve the documents relevant to a user query.
We present \lib, a novel inverted index library implemented in Rust, designed to deliver high performance and efficiency for information retrieval tasks. \lib leverages Rust's safety and performance to provide a robust solution for building and querying inverted indexes, while offering accessible extensibility through its expressive trait system.
While developing \lib, we revisited the inverted-index literature, thereby reproducing numerous prior works using this new test bench.
We evaluated \lib against existing libraries, demonstrating competitive query performance across various datasets and workloads, with speedups of up to $2\times$ over the current state of the art.
Our results indicate that \lib is a promising tool for researchers and practitioners in the field of information retrieval. 
\end{abstract}

\maketitle

\section{Introduction}

The Inverted Index is a foundational data structure in information retrieval systems~\cite{moffat_zobel,ii_comp,tonellotto_book}. Given a collection of documents, the inverted index maps each unique term to the list of documents in which it appears, enabling efficient query processing by rapidly identifying relevant documents based on their content.
Even in modern large-scale contexts, where vector-based retrieval methods are widely used, the inverted index remains a crucial component of hybrid retrieval pipelines as a first-stage candidate generator~\cite{tonellotto_book}.
Top-$k$ retrieval ranks the matching documents by a scoring function, most commonly BM25~\cite{bm25}, and returns the $k$ highest-scoring ones; in modern pipelines these candidates are then re-ranked with more precise learning-to-rank neural models~\cite{tonellotto_book}.

More specifically, inverted indexes also underpin recent approximate nearest-neighbor algorithms for learned sparse retrieval, such as IOQP~\cite{ioqp} and Seismic~\cite{seismic}, the latter being the fastest solution known to date.

Research on inverted indexes centers on two tightly coupled problems: 1) compressing posting lists, so that indexes fit in the memory hierarchy and decode fast, and 2) processing queries efficiently over the compressed data. Below we briefly review the rich literature behind this classic data structure.

\textbf{Posting list compression.} Posting lists are typically gap-en\-coded and then compressed~\cite{ii_comp}. We consider four representative schemes that span the compression/decoding-speed spectrum: Stream VBy\-te~\cite{svb2018}, a SIMD-friendly byte-aligned variant of variable-byte coding~\cite{G8IU}; Binary Interpolative Coding~\cite{bic2000}, which excels on highly clustered lists at the cost of slower decoding; Elias-Fano~\cite{elias1974,fano1971}, which supports constant-time random access and efficient \texttt{nextGEQ} directly on the compressed sequence; and Partitioned Elias-Fano (PEF)~\cite{PEF}, which exploits local clustering by splitting each list into chunks chosen by an $\epsilon$-optimal dynamic-programming partition.

\textbf{Query processing algorithms.} Posting lists can be traversed document-at-a-time (DAAT), term-at-a-time (TAAT), or score-at-a-time (SAAT)~\cite{tonellotto_book}. In this paper we focus on the more popular DAAT algorithms, which process all the posting lists involved in the query simultaneously. Top-$k$ retrieval can be sped up via WAND~\cite{WAND} and MaxScore~\cite{maxscore}, which prune the traversal by maintaining per-list score upper bounds and skipping documents that cannot enter the top-$k$. Block-Max indexes~\cite{blockmax} tighten these bounds by storing a maximum score for each block of the posting list, giving rise to the Block-Max WAND and Block-Max MaxScore variants. The original proposal uses fixed-size blocks~\cite{blockmax}; Mallia \emph{et al.}~\cite{VBMW} later refined the approach with variable-size blocks chosen to minimize the size of the resulting auxiliary structure.

A number of open-source projects continue to push the efficiency of inverted-index implementations in production settings. Among the most widely adopted are Lucene~\cite{apachelucene}, a Java-based search library widely used in industry; Tantivy~\cite{tantivy}, a Rust-based search engine library designed for production use; Vespa~\cite{vespa}, a C++ platform for large-scale search and serving.

In this work, we present \lib, a new efficient inverted index library written in Rust. Our contributions are as follows:
\begin{itemize}
  \item \textbf{A modern, easy-to-use open-source library for inverted index research.} We release \lib as a self-contained Rust library covering the full retrieval stack: index construction, posting-list compression with state-of-the-art encodings --- Elias-Fano~\cite{elias1974,fano1971}, PEF~\cite{PEF}, Binary Interpolative Coding~\cite{bic2000}, and Stream VByte~\cite{svb2018} --- and query processing for Ranked AND/OR, and Top-$k$ retrieval via WAND~\cite{WAND}, MaxScore~\cite{maxscore}, and the Block-Max variants of both~\cite{blockmax}, including variable-size blocks~\cite{VBMW}. The library's trait-based design decouples storage from query logic, making it straightforward to extend, while the absence of external non-Rust dependencies ensures easy adoption via Cargo.
  \item \textbf{Competitive query-time performance.} We evaluate \lib against \dsii and \pisa, two state-of-the-art C++ libraries, on two large web collections. \lib matches or surpasses both competitors on query processing while achieving comparable compression ratios, with speedups of up to $2\times$ compared to \pisa.
  \item \textbf{Reproduction of key results from the literature.} In developing \lib we re-implemented and validated the main compression and query-processing algorithms from the in\-verted-index literature~\cite{moffat_zobel,ii_comp,tonellotto_book}. Specifically, we reproduce the compression and query-time results of Partitioned Elias-Fano indexes~\cite{PEF}, Block-Max WAND with fixed-size blocks~\cite{blockmax}, Block-Max WAND with variable-size blocks~\cite{VBMW}, and the query-length performance analysis of Mackenzie \emph{et al.}~\cite{tutorial_ecir2026}.
\end{itemize}

\noindent\textbf{Resource availability.}
\lib is publicly available on GitHub\footnote{\url{https://github.com/AngeloSav/rise-rs}} under the permissive MIT license.
The repository includes comprehensive documentation of all implemented data structures and algorithms, code examples covering the main use cases, and clear instructions for building, running experiments, and extending the library with other compression algorithms and query processing strategies.

\textbf{Novelty.} While mature C++ libraries such as \dsii and \pisa exist, no equivalent native Rust library covering the full inverted-index stack was previously available. \lib fills this gap, offering a safe, idiomatic alternative that does not sacrifice performance.

\textbf{Utility.} The library is designed to be accessible to researchers with a basic knowledge of Rust and information retrieval: its trait system makes it straightforward to plug in new compressors or query operators without modifying existing code.

\textbf{Impact.} Inverted indexes underpin virtually every text search system, and their study remains an active research area driven by the growing adoption of hybrid sparse-dense retrieval pipelines, as evidenced by recent tutorials on this topic~\cite{tutorial_ecir2026}. We expect \lib to serve as a reproducible test bench for future work on compression algorithms and query processing strategies, and we plan to maintain and extend it as the field evolves.

\section{Library Design}
The Rust programming language has been steadily gaining popularity in recent years due to its combination of safety and performance. Its expressive type system lets us encapsulate behavior through traits, enabling generic code that can be easily extended in the same fashion as many higher-level languages. Performance is not sacrificed: the Rust compiler performs monomorphization and inlining to the same extent as C++ templates. Our main objective in developing \lib was to match the query-time performance of the best existing C++ libraries while retaining this trait-based extensibility.

In \lib, storage and query logic are decoupled: the index exposes the data, while query operators implement retrieval strategies independently of the underlying representation. This separation lets researchers focus on a specific aspect of the retrieval pipeline without touching the rest. The main components, each defined as a trait, are the following:
\begin{itemize}
    \item \texttt{InvertedIndex}: a trait representing the inverted index data structure, providing access to posting lists through its associated \texttt{PostingListIter}.
    \item \texttt{PostingListIter}: a stateful iterator over the posting list of a term; each element is a (docid, frequency) pair.
    \item \texttt{QueryOperator}: a query algorithm that can be executed on any \texttt{InvertedIndex}.
    \item \texttt{DocScorer}: a scoring function used to rank documents by relevance to a query.
\end{itemize}

\begin{figure}[h]
    \begin{center}
    \includegraphics[width=0.9\columnwidth]{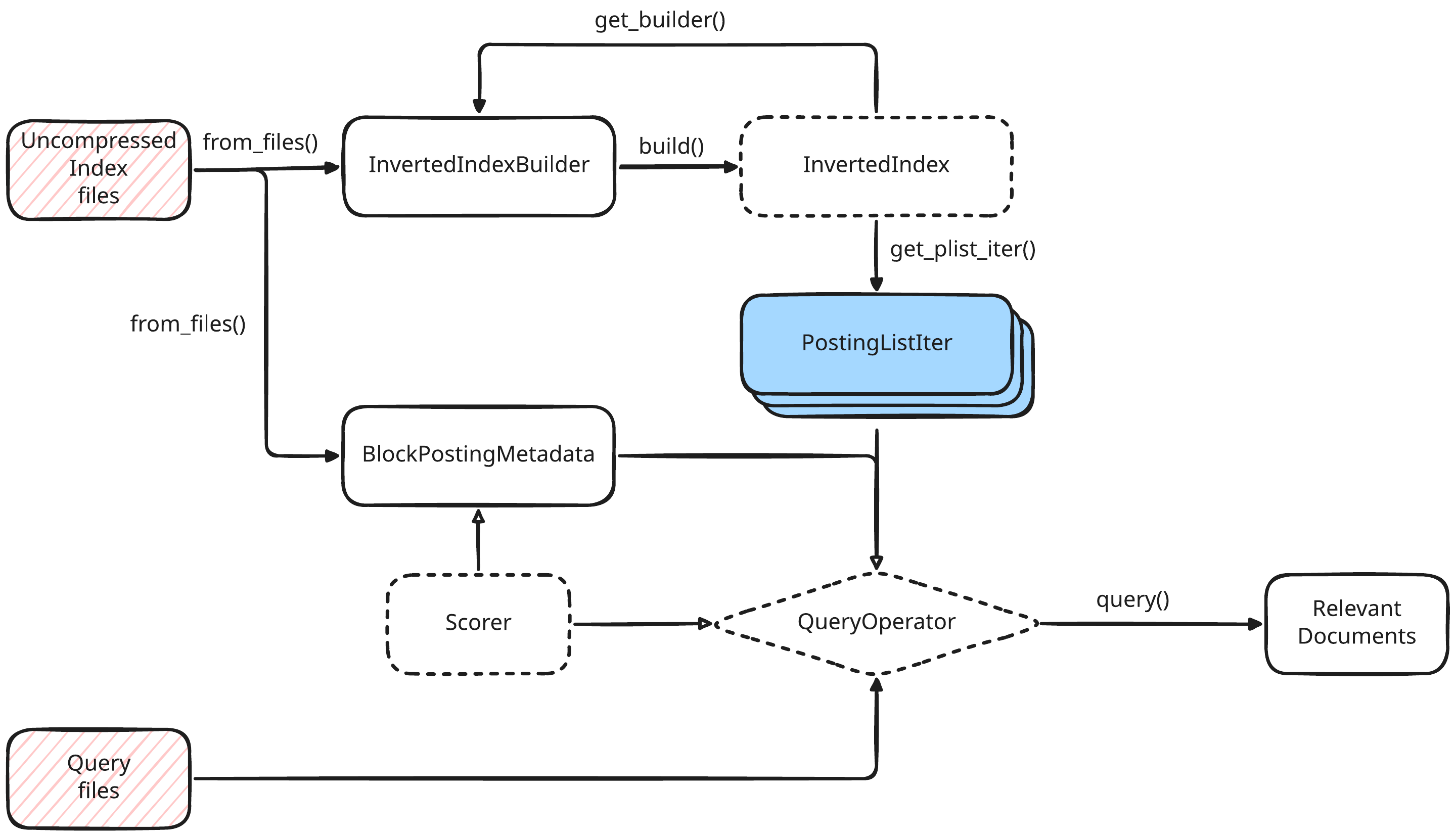}
    \end{center}
    \caption{\label{fig:traits}A high-level overview of the main traits and their interactions in \lib.}
\end{figure}

\Cref{fig:traits} shows a high-level overview of the main traits and their interactions in \lib. Any combination of \texttt{InvertedIndex}, \texttt{DocScorer}, and \texttt{QueryOperator} can be freely composed: for example, a new scoring function requires only implementing \texttt{DocScorer}, and it immediately works with every existing index and query operator.

As a further example of this genericity, we generalized the PEF $\epsilon$-optimal partitioning algorithm so that the optimization algorithm can be applied to any compressor by defining an associated type \texttt{CostWindow} that specifies the cost function.

\section{Experimental Evaluation}

We tested \lib on a variety of datasets and query workloads to evaluate its performance against existing libraries.
The tests have been performed on a machine with an Intel Core Ultra 7 265K 20-core CPU, 8 of which are performance cores with a base frequency of 3.9 GHz. The system is equipped with 128 GB of RAM and is running Ubuntu 25.04 (kernel ver. 6.14.0-37-generic).

\begin{table}[h]
\caption{General information about the datasets used for experimentation}
\label{tab:dataset_info}
\begin{tabular}{@{}lrr@{}}
\toprule
          & \cw      & \cc         \\ \midrule
Documents & 50\,131\,015     & 43\,530\,315     \\
Terms     & 92\,094\,694     & 43\,844\,574     \\
Postings  & 15\,857\,983\,641 & 20\,150\,335\,440 \\ \bottomrule
\end{tabular}
\end{table}

\begin{table*}[!t]
    \centering
    \caption{\label{tab:pef_compression}Index sizes (in bits per integer) of the different compressors implemented in \lib\ on the two datasets, together with the average query time (in milliseconds) of the different query processing strategies.}

\begin{tabular}{llrrrrrrrrr}
\toprule
 &  & \multicolumn{5}{c}{} & \multicolumn{2}{c}{static} & \multicolumn{2}{c}{variable} \\
 &  & BPI & R-AND & R-OR & WAND & MaxScore & BMW & BMMS & BMW & BMMS \\
dataset & index &  &  &  &  &  &  &  &  &  \\
\midrule
\multirow[t]{6}{*}{\cw} & \svb & 11.7 & 7.1 & 72.9 & 9.7 & 6.8 & 3.8 & 4.6 & 3.2 & 3.9 \\
 & \ef & 5.0 & 9.8 & 117.3 & 12.4 & 10.7 & 4.4 & 6.6 & 3.8 & 4.9 \\
 & \upef & 4.0 & 12.3 & 128.8 & 15.2 & 11.0 & 4.8 & 6.3 & 4.0 & 5.2 \\
 & \opt & 3.7 & 10.7 & 123.5 & 13.3 & 10.9 & 4.4 & 6.1 & 3.7 & 4.9 \\
 & \optcomp & 3.6 & 11.5 & 126.9 & 14.3 & 11.4 & 4.6 & 6.3 & 3.8 & 5.1 \\
 & \bic & 3.5 & 35.5 & 239.5 & 45.0 & 18.8 & 16.1 & 12.0 & 10.1 & 10.4 \\
\cline{1-11}
\multirow[t]{6}{*}{\cc} & \svb & 11.2 & 6.8 & 61.8 & 10.1 & 7.2 & 3.7 & 6.7 & 2.9 & 5.4 \\
 & \ef & 4.3 & 9.8 & 98.0 & 13.0 & 11.1 & 3.9 & 7.7 & 3.4 & 6.7 \\
 & \upef & 2.9 & 11.4 & 109.5 & 15.4 & 11.5 & 4.4 & 7.9 & 3.6 & 7.0 \\
 & \opt & 2.6 & 10.0 & 105.9 & 13.7 & 11.5 & 3.8 & 8.0 & 3.2 & 6.9 \\
 & \optcomp & 2.5 & 11.7 & 110.7 & 14.8 & 14.4 & 4.0 & 8.1 & 3.3 & 6.8 \\
 & \bic & 2.5 & 32.8 & 205.0 & 45.8 & 19.6 & 15.1 & 13.7 & 9.7 & 12.4 \\
\cline{1-11}
\bottomrule
\end{tabular}

\end{table*}

The indexes are fully loaded into memory before each experiment, and queries are run in single-threaded mode. We compared \lib\ against two C++ baselines: \dsii~\cite{PEF}, by Ottaviano and Venturini, and \pisa~\cite{PISA}, a fork of \dsii\ that refactors and extends its functionality.

\smallskip
\noindent\textbf{Datasets.}
We tested \lib on two standard datasets: \cw and \cc.
\cw is the ClueWeb 2009 TREC Category B test collection, consisting of about 50 million English Web pages crawled between January and February 2009. \cc is a dataset of news freely available from CommonCrawl \footnote{\url{https://commoncrawl.org/blog/news-dataset-available}}. 
\Cref{tab:dataset_info} provides general information about the two datasets.


\begin{table}
    \centering
    \caption{\label{tab:rgb_compression}Index sizes (in GiB) of different compressors on the two datasets, with dataset sorted by URLs and with recursive graph bisection.}
    \begin{tabular}{l ccc ccc}
\toprule
& \multicolumn{3}{c}{\cw} & \multicolumn{3}{c}{\cc} \\
\cmidrule(lr){2-4} \cmidrule(lr){5-7}
index & Random & URL & RGB & Random & URL & RGB \\
\midrule
\svb  & 44.68 & 43.29 & 43.00 & 54.09 & 53.85 & 52.80 \\
\upef & 19.10 & 16.63 & 14.96 & 20.31 & 19.74 & 13.50 \\
\opt  & 18.44 & 15.24 & 13.74 & 19.45 & 18.86 & 12.02 \\
\bic  & 19.18 & 14.24 & 12.98 & 20.26 & 18.72 & 11.65 \\
\bottomrule
\end{tabular}
\end{table}

\smallskip
\noindent\textbf{Compression and Query Processing.}
Many compression techniques with different space-time tradeoffs have been proposed for inverted indexes~\cite{ii_comp}. In \lib, we reimplemented several of them from scratch to keep the library self-contained: Elias-Fano~\cite{elias1974,fano1971} (\ef), Partitioned Elias-Fano~\cite{PEF} in its uniform (\upef) and variable (\opt) partitioning variants, Binary Interpolative coding~\cite{bic2000} (\bic), and StreamVByte~\cite{svb2018} (\svb). The original Partitioned Elias-Fano~\cite{PEF} uses three encoding methods: Elias-Fano, characteristic vector, and all-ones sequence. We also implemented a new variant called \optcomp\ with a fourth option, which handles almost-full sequences by storing their complement. As we will see, this brings the compression of \optcomp\ on par with Binary Interpolative Coding, the most space-efficient compressor in our comparison, while retaining the time efficiency of Partitioned Elias-Fano.

We first evaluated the efficiency of the compressors on our datasets under three document orderings: a random permutation, the original URL ordering, and the assignment produced by recursive graph bisection~\cite{rgb,enhanced_graph_bisection}. URL ordering is known to be a good heuristic when such identifiers are available~\cite{docid_sorting}, particularly on Web crawls. As shown in \Cref{tab:rgb_compression}, random ordering is uniformly the worst across all gap-based compressors, while recursive graph bisection always produces the smallest indexes. The gain of RGB over URL ordering depends strongly on the dataset: on \cw, where URLs already cluster related documents, RGB shrinks the \opt\ index by about $10\%$ (from $15.24$ to $13.74$ GiB); on \cc, where URLs carry much less structural signal, the same reordering shrinks the \opt\ index by roughly $36\%$ (from $18.86$ to $12.02$ GiB), with similar gains for \upef\ and \bic. \svb\ benefits only marginally, as its byte-aligned encoding is largely insensitive to gap distributions. The improvement comes with no measurable loss in query performance, and for this reason all subsequent experiments use indexes reordered with recursive graph bisection.

\begin{table}[t]
    \centering
    \caption{\label{tab:blockmax_table} Query performance across the three libraries on the two datasets, restricted to three representative compressors (\svb, \opt, \bic). For \dsii, we use its \texttt{varint-G8IU} encoding when comparing against the StreamVByte indexes of the other libraries. The query times are measured in milliseconds.}
    \setlength{\tabcolsep}{4pt} 
\begin{tabular}{lllrrr HH rr}
\toprule
 &  &  & \multicolumn{3}{c}{} & \multicolumn{2}{c}{} & \multicolumn{2}{c}{variable} \\
 &  &  & R-OR & WAND & MaxScore & BMW & BMMS & BMW & BMMS \\
 & index & lib &  &  &  &  &  &  &  \\
\midrule
\multirow[c]{9}{*}{\rotatebox[origin=c]{90}{\cw}} & \multirow[t]{3}{*}{\svb} & \lib & 72.9 & 9.7 & 6.8 & 3.8 & 4.6 & 3.2 & 3.9 \\
 &  & \pisa & 110.9 & 9.5 & 5.9 & 5.0 & 8.7 & 3.7 & 8.5 \\
 &  & \dsii & 64.2 & 9.1 & 5.2 & 5.1 & - & 3.6 & - \\
\cmidrule{2-10}
 & \multirow[t]{3}{*}{\opt} & \lib & 123.5 & 13.3 & 10.9 & 4.4 & 6.1 & 3.7 & 4.9 \\
 &  & \pisa & 160.4 & 12.8 & 10.4 & 5.5 & 11.5 & 4.0 & 10.8 \\
 &  & \dsii & 114.8 & 12.3 & 8.9 & 5.3 & - & 3.7 & - \\
\cmidrule{2-10}
 & \multirow[t]{3}{*}{\bic} & \lib & 239.5 & 45.0 & 18.8 & 16.1 & 12.0 & 10.1 & 10.4 \\
 &  & \pisa & 312.8 & 52.0 & 19.6 & 19.6 & 20.6 & 12.4 & 19.5 \\
 &  & \dsii & 236.6 & 46.6 & 17.2 & 17.4 & - & 10.1 & - \\
\midrule 
\multirow[c]{9}{*}{\rotatebox[origin=c]{90}{\cc}} & \multirow[t]{3}{*}{\svb} & \lib & 61.8 & 10.1 & 7.2 & 3.7 & 6.7 & 2.9 & 5.4 \\
 &  & \pisa & 94.2 & 11.2 & 7.0 & 4.0 & 10.0 & 3.4 & 9.7 \\
 &  & \dsii & 51.5 & 9.8 & 5.6 & 4.1 & - & 3.3 & - \\
\cmidrule{2-10}
 & \multirow[t]{3}{*}{\opt} & \lib & 105.9 & 13.7 & 11.5 & 3.8 & 8.0 & 3.2 & 6.9 \\
 &  & \pisa & 132.5 & 14.3 & 12.3 & 4.2 & 15.3 & 3.5 & 12.6 \\
 &  & \dsii & 92.3 & 13.2 & 9.8 & 4.1 & - & 3.4 & - \\
\cmidrule{2-10}
 & \multirow[t]{3}{*}{\bic} & \lib & 205.0 & 45.8 & 19.6 & 15.1 & 13.7 & 9.7 & 12.4 \\
 &  & \pisa & 251.9 & 49.6 & 21.0 & 15.0 & 22.4 & 10.9 & 21.7 \\
 &  & \dsii & 189.9 & 45.1 & 18.1 & 13.7 & - & 9.5 & - \\
\bottomrule
\end{tabular}
\end{table}

On top of these compressors, \lib implements the main query processing strategies in use today: ranked AND (R-AND), ranked OR (R-OR), WAND, and MaxScore, together with their block-max variants Block-Max WAND (BMW) and Block-Max MaxScore (BMMS), in both static and variable-block partitioning. We create the block-max indexes using the approaches described by Ding and Suel~\cite{blockmax} for static partitioning, and by Mallia \emph{et al.}~\cite{VBMW} for variable partitioning. Since our aim is to evaluate the cost of the algorithms themselves, we store the block-max indexes in their uncompressed form, as it provides the fastest access times.

In \lib\ we implement a different version of BMMS compared to \pisa, adapted from Apache Lucene's block-based scoring strategy. We use a double windowing approach: in the outer window, the block maxima drive the MaxScore essential/non-essential split; in an inner window of fixed size, we score all documents without consulting the top-$k$ heap, and update it only at the end of the window. This amortizes the cost of updating the heap across many documents, making queries more efficient when the documents to evaluate are dense (e.g.\ long or common-term queries).

We tuned the block size for the static partitioning approach and the parameter $\lambda$ for the variable partitioning approach, which represents the fixed cost associated with creating a new block. The best performance is achieved with a static block size of $128$ or $\lambda = 12$ for variable-block partitioning, consistent with the defaults used by \pisa\ and \dsii. We adopt these values in all the remaining experiments.

\Cref{tab:pef_compression} reports the index size in bits per integer (BPI) and the average query time (in ms) for every compressor/algorithm combination in \lib. \svb\ is the fastest but uses more than double the BPI of the PEF family. \optcomp\ closes the space gap with \bic\ — the smallest but slowest by a wide margin — while retaining the query times of the rest of the PEF family, a direct effect of its fourth encoding method for almost-full sequences. Across query strategies, variable-block partitioning is consistently faster than static partitioning for both BMW and BMMS, and variable-partition BMW achieves the overall best query time on both datasets.


\smallskip
\noindent\textbf{Impact of query length.}
We evaluate the impact of query length on the performance of the different query algorithms. We use the \cc dataset and we construct indexes using StreamVByte, as it is the one that offers the best performance in terms of query time. We then divide our query workload into different batches, where batch $i$ contains all the queries with exactly $i$ terms. The performance of all algorithms degrades as the number of terms increases, but the slopes differ sharply. Block-max algorithms dominate on short queries, where their upper-bound skipping is most effective: at $2$ terms, BMW takes about $1.5$ ms against roughly $6$ ms for MaxScore, and BMW is always faster than standard WAND. MaxScore is by far the flattest curve, growing only from about $4.5$ ms at $1$ term to $17.5$ ms at $9{+}$ terms. It overtakes the block-max algorithms past $5{-}6$ terms; in the longest bucket, MaxScore reaches roughly $17.5$ ms against $32$ ms for BMW. Block-max algorithms are thus preferred when queries are short and keyword-matching is the primary goal, while for longer queries or query expansion MaxScore's lower overhead makes it the better choice. These results are in line with the ones reported by Mackenzie \emph{et al.}~\cite{tutorial_ecir2026}, and we consider them to be reproduced by our experiments.

\smallskip
\noindent\textbf{Cross-library comparison.}
We now compare the query performance of \lib\ against \dsii\ and \pisa\ on the same indexes, restricting the comparison to three representative compressors: \svb\ (fastest), \opt\ (best space-time tradeoff in the PEF family), and \bic\ (most space-efficient). We adopt a fixed block size of $128$ and $\lambda = 12$ across all libraries; \dsii\ does not implement BMMS, so those entries are missing in \Cref{tab:blockmax_table}.

For the standard union algorithms (Ranked OR, WAND, MaxScore), \dsii\ remains the best performer, especially when the number of documents is large. This is the case especially in Ranked OR, where no document skipping is performed. \pisa\ is the slowest library, likely because its heavier use of C++ abstractions makes inlining harder for the compiler. By contrast, \lib's use of traits and immutable references yields better-compiled code, resulting in query-time performance much closer to \dsii\ than to \pisa, despite using higher-level language features.

For block-max algorithms, \lib\ is reliably faster than both \pisa\ and \dsii\ on BMW, in both static and variable partitioning. Furthermore, the BMMS implementation in \lib\ is much faster than the one in \pisa, with a speedup of up to $2\times$ in the variable-block version of the algorithm. Overall, \lib\ matches or improves on the state of the art on the algorithms that matter the most in practice, while keeping the engineering benefits of a high-level, memory-safe implementation.

\section{Conclusion}
We introduced \lib, a modern inverted index library written in Rust, designed to provide a high-performance native implementation for efficient keyword retrieval. We have shown that \lib offers query performance that is often faster than state-of-the-art libraries, while providing a safe and modular codebase. We believe that \lib can be a valuable tool for researchers and practitioners in the field of information retrieval.
Possible further work includes the implementation of other compression techniques, such as PForDelta~\cite{pfor} and QMX~\cite{QMX}, and Roaring bitmaps~\cite{roaring}, block-based scoring, recently adopted in many commercial search engines, and specialized support for learned sparse retrieval datasets. We plan to keep extending and maintaining \lib\ in the future, to ensure its continued relevance and performance.

\bibliographystyle{ACM-Reference-Format}
\bibliography{fiir_paper} 

\end{document}